\documentclass[journal,11pt,onecolumn]{IEEEtran}

\usepackage{fancyhdr, epsfig, epsf, amsthm, amsmath, amssymb, amsfonts}
\usepackage[noadjust]{cite}
\usepackage[utf8]{inputenc}
\usepackage{stmaryrd }
 \usepackage{tikz}
\usepackage[T1]{fontenc}
\usepackage{boisik}

\usepackage{dsfont,enumerate,enumitem,comment,url}
\usepackage{multirow,booktabs,bigstrut,longtable,array}
\usepackage{colortbl,pdflscape,tabu,threeparttable,threeparttablex}
\usepackage[normalem]{ulem}
\usepackage{graphicx,color,rotating,soul}
\usepackage{makecell,xcolor,pbox,framed}
\usepackage{wrapfig,setspace,subfigure,float}
\usepackage[most]{tcolorbox}    
\tcbuselibrary{breakable} 

\usepackage{hyperref}

\colorlet{shadecolor}{gray!20}

\usepackage[a4paper,left=20mm,right=20mm,top=15mm,bottom=15mm]{geometry}

\usepackage{fancyhdr}

\RenewDocumentCommand {\maketitle} {} {%
    \newgeometry{left=-16.5pt, top=-13.25pt, right=0pt, bottom=0pt}
    \begin{tikzpicture}[remember picture, overlay]
        \node[anchor=north west, inner sep=0, outer sep=0] at ($(current page.north west)$) {
            \includegraphics{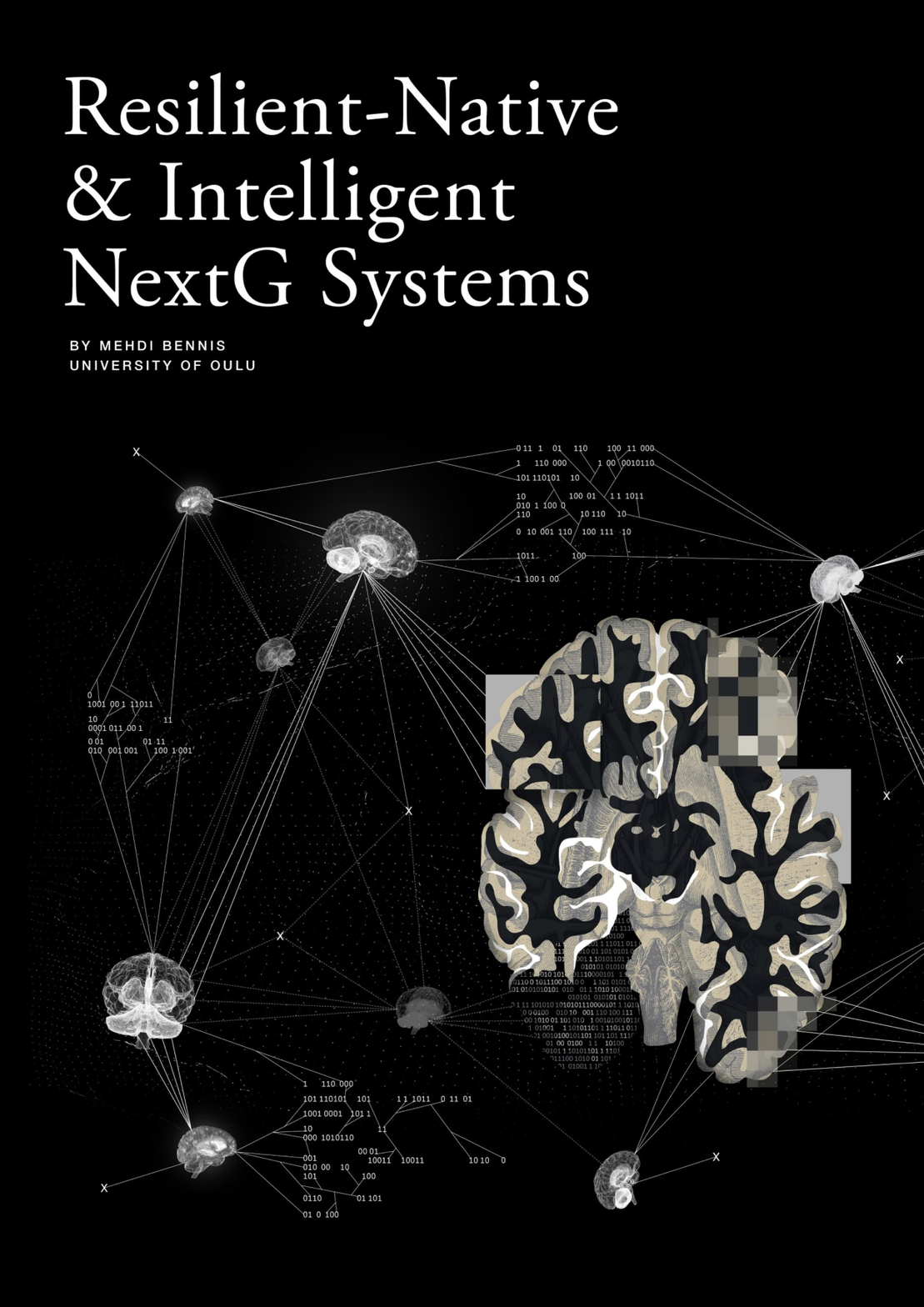}
        };
    \end{tikzpicture}
    \restoregeometry{}
}
\usetikzlibrary{calc} 

\begin{document}
\maketitle


\DeclareRobustCommand{\hlcyan}[1]{{\sethlcolor{green}\hl{#1}}}
\DeclareRobustCommand{\hlblue}[1]{{\sethlcolor{cyan}\hl{#1}}}

\begin{abstract}
Just like power, water and transportation systems, wireless networks are
a crucial societal infrastructure. As natural and human-induced disruptions continue to grow,
wireless networks must be resilient to unforeseen events, able to withstand and recover from
unexpected adverse conditions, shocks, unmodeled disturbances and cascading failures. Despite its
critical importance, resilience remains an elusive concept, with its mathematical foundations still
underdeveloped. Unlike robustness and reliability, resilience is premised on the fact that disruptions
will inevitably happen. Resilience, in terms of elasticity, focuses on the ability to bounce back to
favorable states, while resilience as plasticity involves agents (or networks) that can flexibly expand
their states, hypotheses and course of actions, by transforming through real-time adaptation and
reconfiguration. This constant situational awareness and vigilance of adapting world models and
counterfactually reasoning about potential system failures and the corresponding best responses,
is a core aspect of resilience. This article seeks to first define resilience and disambiguate it from
reliability and robustness, before delving into  the mathematics of resilience.
Finally, the article concludes by presenting nuanced metrics and discussing trade-offs tailored to
the unique characteristics of network resilience. 
\end{abstract}

 \section{Resilience: What and How?}
\begin{wrapfigure}{R}{7.5cm}\centering \vspace{-1pt}
 \includegraphics[width=7.5cm]{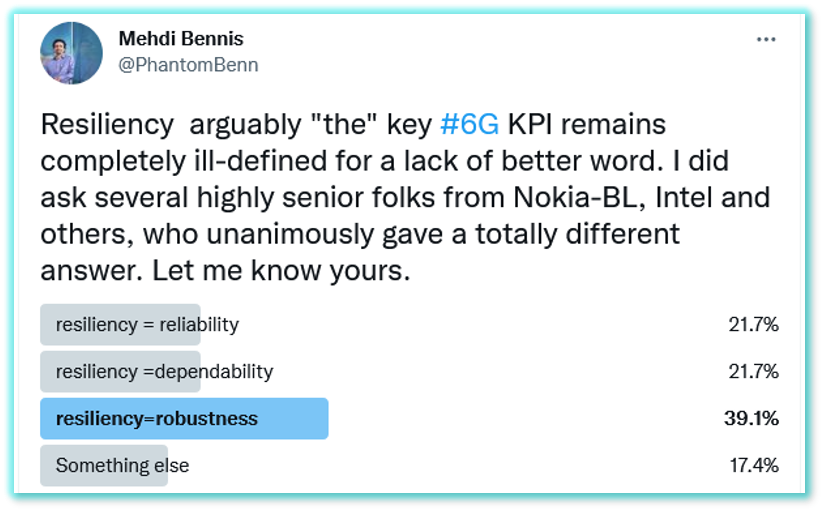} 
\caption{\footnotesize  A poll on resilience (source: Twitter/X).} \vspace{-8pt}
	\label{first}
\end{wrapfigure}

Resilience, arguably one of the key 6G performance indicators
(KPIs) has been discussed in various fora, panels, and
articles \cite{ ieeespectrum, resil, saad2019vision, khaloopour2024resiliencebydesignconcepts6gcommunication, reifert2022comeback, vesterby_2022,9963527}.  Yet to this day, resilience remains an elusive and ill-defined concept where anything goes (see  Fig.~\ref{first} for a recent poll).   
Despite its crucial and far-reaching importance, the  mathematical foundations of resilience are sorely lacking \cite{ieeespectrum}.  Resilience is often equated with reliability or robustness causing  confusion. On the one hand, robustness\footnote{Robust control addresses known uncertainties but is oblivious to unforeseen ones—robust, yet inherently fragile.}  is the ability to withstand adverse conditions and accommodate known uncertainties, such as measurement disturbances and sensory/actuators faults. Robustness can be modeled as a minmax optimization problem, or a zero-sum game using risk-sensitive objective functions.   On the other hand, reliability (one of the distinguishing features of 5G) deals with the statistics of extreme/rare events, in which the goal is to characterize  and tame the tail distribution of performance metrics, such as transmission rate and latency \cite{8472907,7529226}. For instance, a  link-level reliability of 99$\%$ implies  at most one out of  $100$ packet losses occurs within a pre-defined time period and given known uncertainty (known unknowns). 
 However, these definitions fall short in scenarios  requiring performance guarantees in the presence of consecutive losses over an extended period, nor (and quite crucially) do they provide any assurance of desired functionalities
under unexpected or unknown stressors (unknown unknowns). In sharp contrast to both robustness and reliability, resilience assumes unforeseen events will occur, warranting the ability to  withstand  stressors (resist), recover  and adapt in real-time. 
 Stressors  can be known or unknown, ranging from node/network/component failure, misinformation, compromised sensor/node, adversarial  attacks, jamming, link/network outages and other    environmental factors.

\vspace{+1mm}
While both robustness and resilience address the impact
of stressors on system functionality, they differ in their objectives. Robustness  optimizes for  worst-case scenarios and focuses on maintaining performance levels over a
limited and known range of stressors. In contrast to robustness,  resilience does not aim at entirely preventing failures and is rooted in quick
recovery from disruptions and adaptation (plasticity)  under unknown stressors. 
This fundamental distinction arises from the capacity for online adaptation or
configuration inherent in resilience, in contrast to offline design 
principles associated with robust approaches. Consequently, robust designs often require higher resource consumption
to withstand known stressors (e.g., via redundancy). In contrast, resilient
designs  adapt to new types of stressors, recover and sustain  functionality as long as possible, while (possibly) undergoing structural transformation. What is more? there exists an interesting and unexplored robustness-resilience-efficiency tradeoff. Indeed,  a network can be robust  against $N$-node failure, yet fail for larger number of nodes, calling for resilience.  As the scale of potential failures increases, preparing for all conceivable N-node or N-link failure becomes impractical, hence resilience becomes a must-have, where the network readies itself to confront unforeseen events.   This underscores the tradeoff wherein  robustness, as a precursor to resilience,  contributes to resilience while resilience can mitigate  events that cannot be handled through robustness. Simply put, when unforeseen events  lie beyond the system's prior knowledge, resilience is the only option.  \begin{figure}
	  \begin{center}
              \includegraphics[width=\linewidth]{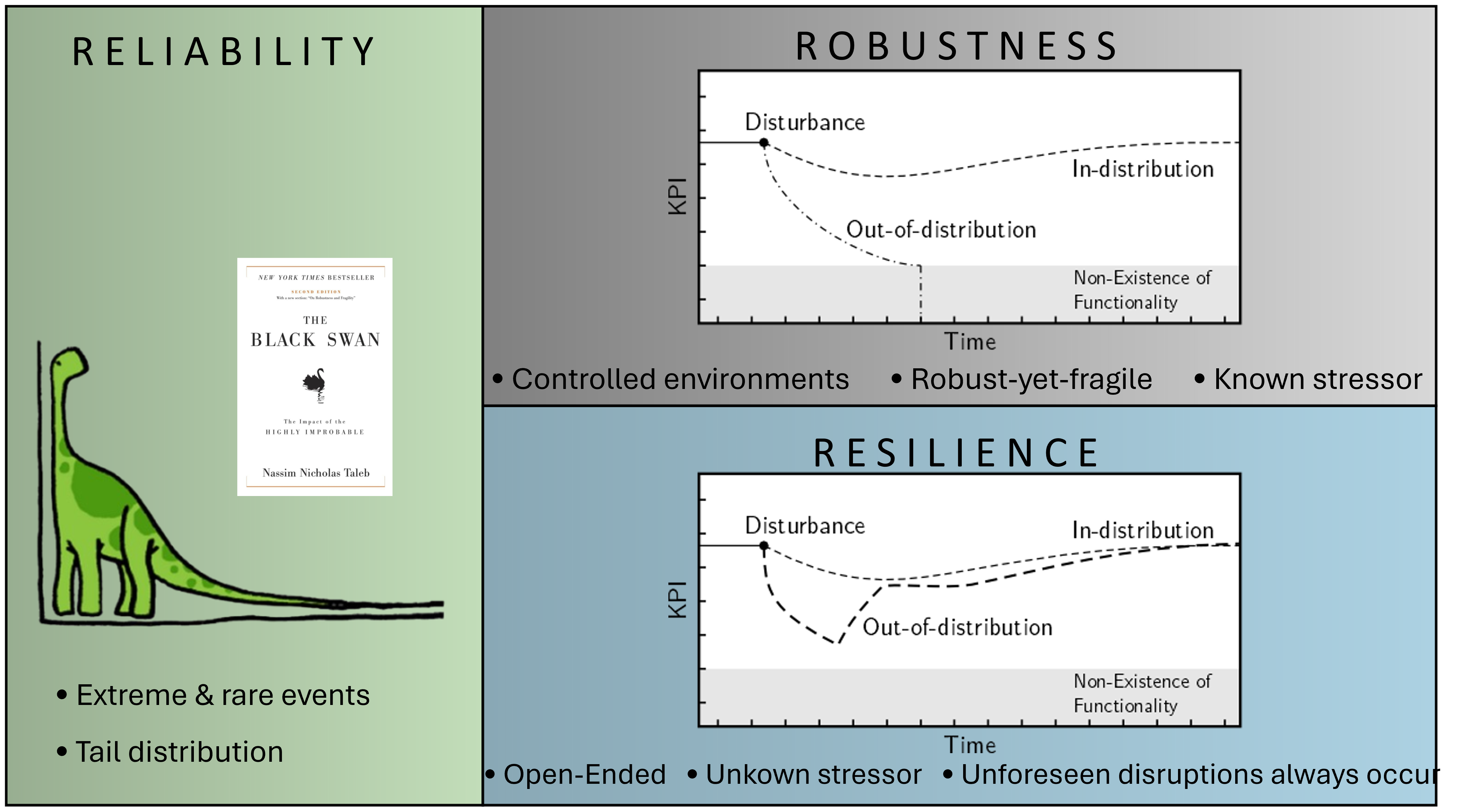} 

	  \end{center} \caption{\footnotesize  The three R's of reliability, robustness and resilience.} 
	\label{dino}
\end{figure} 
\vspace{+1mm}

Resilience is found at various microscopic and macroscopic scales, ranging from biological brains, animals (e.g., bird flocks, ants colony) to human-made power grids, cities and  other infrastructure systems.   Taking an inspiration from the brain, (loosely) seen as  networked, heterogeneous, hierarchical and modular control-loops (or agents) that  sense their environments, perceive, abstract,  predict/infer, communicate  and plan accordingly, resilience can be seen through the lens of elasticity and plasticity. While elasticity  simply means returning to preferred states after disruption, plasticity requires planning, structural reorganization/ reconfigurability andtransformation.  In short,   while reliability aims at preventing failures and breakdowns, resilient nodes (networks)  recover, maintain a desired functionality  and adapt  under unexpected events.    
\subsection{Abridged State-of-the-art}
Over the years, countless articles, surveys, and books have been written on resilience across various fields such as ecology, social networks, engineering, robotics, cyber-physical systems, and very recently 6G (e.g., \cite{ Royce, mitre, khaloopour2024resiliencebydesignconcepts6gcommunication,STERBENZ20101245, resil, reifert2024resiliencecriticalitybrothersarms,weinberger2025acceleratedrecoveryrisdesigning}, to name just a few).  Within 6G, resilience has gained significant attention, as evidenced by the NSF-funded Resilient $\&$ Intelligent NextG Systems (RINGS I-II) programs \cite{WinNT}. Despite this growing interest, resilience remains vaguely defined, often equated with robustness, and lacks concrete metrics. Furthermore, resilience is frequently conflated with robustness and reliability, emphasizing the need for a rigorous and much-needed mathematical formalism for defining,  quantifying and optimizing for resilience.

\vspace{-2mm}
\subsection{Multidisciplinary Impact}
Given the increasing severity and frequency of unforeseen events\footnote{Such as the  2024 Boeing crash and recent 2025 Spanish power outage.} and attacks, combined with the disaggregated, distributed, and virtualized nature of networks, resilience has become a necessity rather than an option. Within 6G, resilience offers the potential to redefine the current vision(s) and its requirements, which to date have been an incremental evolution of 5G \cite{ieeespectrum}, \cite{saad2019vision}. Unlike reliability and robustness, resilience is not about entirely preventing failures through resource overprovisioning (more base station deployments or pursuing extremely low packet error rates) and addressing expected uncertainties. Instead, resilient nodes and networks must adapt to unforeseen events and transform in an efficient manner, maintaining resilience without sacrificing efficiency. Within ML/AI, resilience plays a transformative role, particularly in out-of-distribution (OOD) generalization, safety,   formal specification and verification of models, especially for mission/safety-critical applications where  statistical models fall short.

\section{Fundamental Questions and Mathematics of resilience}
Studying resilience through first principles hinges on addressing several fundamental questions, sitting at the intersection of many disciplines. Principles of resilience are rooted in abstraction,
anticipation, adaptation/transformation at node/system-level, compositionality and emergence via communication and interaction. Namely, for situational awareness and active preparedness,
agents must learn abstractions  and world models from 
multimodal sensorimotor signals, plan ahead,
counterfactually reason and adapt accordingly via communication. Moreover, the  interconnected nature of nodes and subsystems warrants an investigation  of the compositional nature of resilience. The following \textit{four} research questions   ground resilience in a nuanced and technical treatment.

\subsection{Abstraction, Anticipation, Adaptation ($A^3$)}

\begin{tcolorbox}[colback=yellow!10, colframe=white!20!black, boxrule=1.9pt, boxsep=2pt,left=2pt,right=2pt,top=2pt,bottom=2pt]
    \noindent \textsf{Q1}: For situational awareness and active preparedness, how do agents learn abstractions  from  (multimodal) sensory signals? How long can a node (or network) resist and function by preserving its internal  geometrical/topological structures? After disruption,  how to  speed up recoverability  and minimize timelapses during which the functionality and structure are
compromised? 
    \end{tcolorbox}

Addressing Q1 \begin{wrapfigure}{R}{5cm}\centering \vspace{-2pt}
            \includegraphics[width=\linewidth]{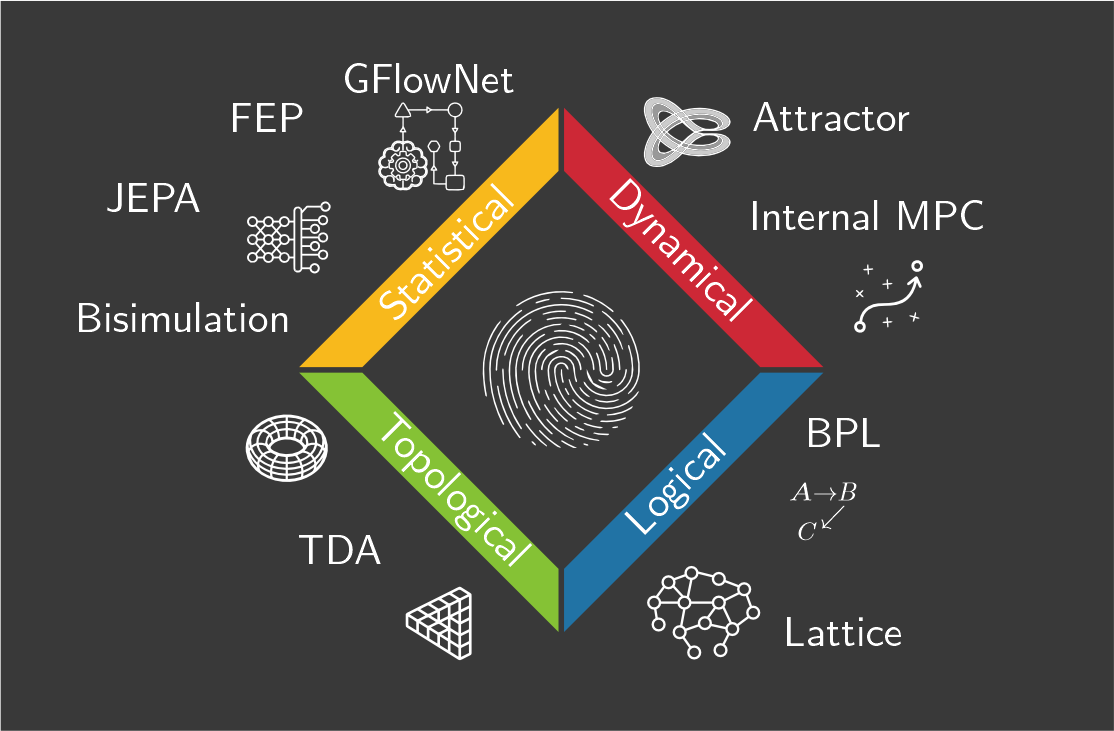} 

	\caption{\footnotesize Abstractions along the statistical, logical, dynamical and topological continuum.}\vspace{-1pt}
	\label{fig4}
\end{wrapfigure} requires the design of agents (or networks of agents) that can sense, monitor, and continuously update their internal representations of their external environment. This adaptive capability can be modeled through various frameworks, such as the Free Energy Principle (FEP) \cite{FRISTON200670, WAFR}, which formulates perception and action through variational Bayesian inference. A core requirement of these approaches is the ability to infer internal structured representations (abstractions) of the environment from internal feedback, in which different sensory data give rise to a plethora of abstractions (listed in Fig. \ref{fig4}). For hierarchical-type data, such as multi-resolution sensory signals, algebraic structures like \textit{lattices} offer a formal mechanism to represent equivalence relations and refinement operations,  enabling efficient structural compression and fast inference. From a dynamical systems perspective, abstraction is associated with low-dimensional attractors, or attractor-like structures. These attractors can be learned using methods such as the Generative Flow Networks (GFlowNets), which learn to sample a (diverse) set of state trajectories, such that the distribution of terminal states is proportional to an unnormalized target density \cite{bengio2023gflownet}. Subsequently,   these local attractors can be flexibly reused and recomposed to solve specific downstream tasks. In the context of formal logic, abstractions correspond to probabilistic programs, which are modular constructs that allow agents to represent and compose concepts hierarchically \cite{doi:10.1126/science.aab3050}. Unlike statistical inference techniques (e.g., FEP), program inference is tantamount to  learning compressed mathematical theories (minimal programs) that best explain observed sensorimotor data, minimizing Kolmogorov complexity (minimum description length).

Resilience can manifest through three primary responses to environmental perturbations or stressors: (i)   \textit{Resistance}, where the agent maintains its state despite external changes; (ii) 
    Elasticity, wherein the agent returns to a prior stable state following disruption;  and (iii) 
    Plasticity, which involves structural adaptation through the exploration of new state spaces and/or reconfiguration of internal models or network topologies.   Unlike elasticity, plasticity necessitates  learning internal world models that support abstraction, active preparedness via counterfactual reasoning,  strategic planning, and decision-making. These world models allow agents to simulate possible future disruptions and proactively prepare the  best responses to a potential failure or disruption.

World models have been extensively utilized in model-predictive control, where models of system dynamics inform control decisions, and in model-based reinforcement learning, where predictive models of state transitions guide decision-making \cite{DBLP:journals/corr/abs-1803-10122}. Recent developments, such as the Joint Embedding Predictive Architecture (JEPA) \cite{lecun2022path}, have renewed interest in learning latent representations that support long-range planning and predictive reasoning. Beyond statistical methods, logical and symbolic approaches treat planning as a form of automated theorem proving, in which an inference module proposes sequential steps to validate the correctness of a candidate solution relative to a formal model of the world. For agents to be resilient across diverse and dynamic environments, they must leverage a collection of hierarchical and multimodal models. These models may span different syntactic  representations (e.g., probabilistic, geometric, or symbolic), and must support cross-modal knowledge transfer, such as integrating visual and linguistic information streams. Given the inherent epistemic uncertainty of real-world environments, resilient agents must also exhibit a form of epistemic curiosity, by actively seeking information to reduce uncertainty and improve internal models for enhanced decision-making. 
In summary, resilient agents  must be endowed with situational awareness and active preparedness, which requires a synergy of abstraction, (compositional) reasoning using predictive (multimodal) world models and adapting to novel environments.

\vspace{1mm}
\subsection{Algebraic Compositionality  } 

\begin{tcolorbox}[colback=yellow!10, colframe=white!20!black, boxrule=1.9pt, boxsep=2pt,left=2pt,right=2pt,top=2pt,bottom=2pt]
    \noindent \textsf{Q2}:   Owing to the interconnected nature of components, nodes and subsystems, how to algebraically compose them to emerge a resilient system?  How do these distributed algebraic structures, each with distinct syntax and semantics, interact and coordinate effectively to solve a plethora of tasks?    
    \end{tcolorbox}

\begin{wrapfigure}{R}{8cm}\centering \vspace{-1pt}
        \includegraphics[width=\linewidth]{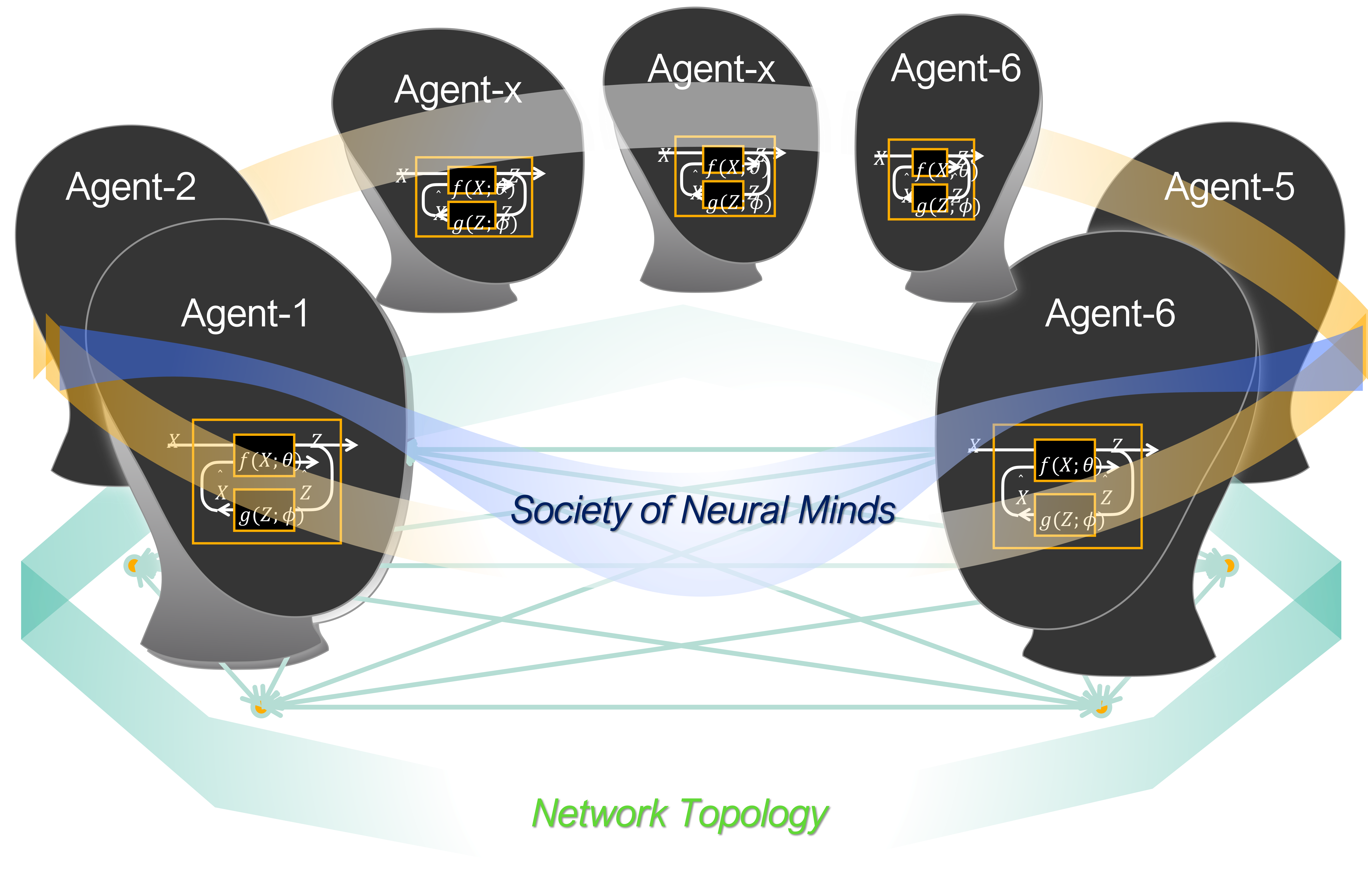} 

	\caption{\footnotesize  Network resilience via  the (judicious) composition of the semantics of active inference agents. } \vspace{-2mm}
	\label{compose}
\end{wrapfigure}

Humans possess a remarkable cognitive capacity to form high-level relational representations (or abstractions) and  combine them in flexible and novel ways to adapt and generalize rapidly in dynamic environments. The algebraic composition of concepts, control loops, and subsystems plays a foundational role in a wide array of  tasks, including reasoning, planning, communication and control (see Fig. \ref{compose} for an illustration). Given the distributed, interconnected, and multiscale nature of agents/subsystems, a central challenge is to understand how resilience emerges and scales as a function of network size, topology, connectivity, and dynamics. As a matter of fact, the resilience of individual subsystems does not necessarily imply the resilience of the entire system. Conversely, globally resilient systems can emerge from non-resilient subsystems \cite{bowers2017team}. This asymmetry mandates a systematic and compositional framework for studying resilience at both local and global spatiotemporal scales.

Algebraic compositionality provides such a framework, enabling the construction of resilient networks from non-resilient nodes (or components), while allowing for  comparisons between networks in terms of resilience—e.g., determining whether one network is more resilient than another, or whether a node exhibits local but not global resilience. For a single agent, algebraic compositionality enables the integration and fusion of distributed internal structures, such as multimodal sensory inputs represented in diverse syntactic or semantic forms. Fusing the semantics of these multimodal sensory signals can be formalized using sheaf theory, which captures how local representations combine into coherent global structures \cite{bredon_sheaf_1997}. Sheaf theory is particularly useful for reconciling inconsistencies between an agent’s multimodal world models—its internal abstractions and beliefs—and for facilitating belief alignment across multiple agents. This alignment supports cross-modal generation, information transfer and predictive modeling of others’ actions and intentions \cite{issaid2025tacklingfeaturesampleheterogeneity}. Furthermore, compositionality allows reasoning   agents to simulate hypothetical actions by combining internal concepts, an essential capacity for adaptability and resilience, as emphasized in the discussion of Q1. What is more? as articulated in the  grand vision in \cite{bennis2025semanticcommunicationmeets2}, algebraic principles  underpin semantic communication, where meaningful and timely information exchange between agents is achieved by composing their internal information spaces (conceptualized as a sheaf of world models) for mutual predictability and understanding.  An analogous perspective is  found in enactivist and dynamical system approaches, whereby each agent is modeled as an information transition system (ITS), and the agent’s internal states define a transition system over its information space \cite{WAFR}. Communication in this context is  the alignment or composition of world models to ensure shared predictions and coordinated behavior. For network resilience, understanding  the inherent trade-offs between information redundancy, diversity, and complementarity is fundamental, where agents must balance between reducing redundancy (multiple components serving the same function) and maximising  plasticity/adaptability (components performing various functions depending on context). 

\subsection{Formal Verification and Multi-agent Logic  }

\begin{tcolorbox}[colback=yellow!10, colframe=white!20!black, boxrule=1.9pt, boxsep=2pt,left=2pt,right=2pt,top=2pt,bottom=2pt]
    \noindent \textsf{Q3}:  As networks are becoming disaggregated, distributed, and virtualized, how to certify and verify the functioning, safety, interoperability, and explainability of models and protocols? For mission/safety-critical applications, how do formal verification methods and   logical reasoning provide a rigorous modeling and optimization framework for resilience, beyond statistical methods?
    \end{tcolorbox}

As AI systems are increasingly deployed in safety/mission-critical applications, relying on probabilistic approaches is ill-suited. Formal verification methods and logic (temporal, epistemic, modal, etc.) are necessary for certifying and explaining the proper functioning of  models, communication protocols and networks, as well as ascertaining resilience under belief manipulation and  misleading information.    For instance, signal temporal logic (STL)  provides a formal specification language for capturing temporal dynamics (e.g., of a real-valued sensory signal $\zeta$) and STL formulas $\phi$ are defined recursively according to  some  grammar,   admitting a quantitative semantics given by a real-valued function $\rho$.  The $\rho$-value, called the robustness satisfaction value (RSV), can be interpreted as the extent to which the signal $\zeta$ satisfies a formula $\phi$ at time t. Its absolute value can be viewed as
the distance of $\zeta$ from the set of trajectories satisfying or violating  $\phi$, with positive values indicating satisfaction and negative values indicating violation.  Importantly, these  quantitative semantics of the logical specifications enable the incorporation of these requirements into an actionable optimization framework for solving specific tasks \cite{chen2023stl}, \cite{GirgisSTL}.  
Specifically, in the context of resilience,  STL-specifications provide a rich language to describe a resilient node/subsystem in terms of: (i) recoverability (given an STL specification, a signal must recover from a violation  within a  pre-defined time period); (ii) durability (extent to which it can maintain its functionality for at least a pre-defined duration).  In terms of formal guarantees,  the STL-based resilience specification allows  to prove the soundness and completeness of its semantics, and  STL quantitative semantics in terms of recoverability-durability pairs provide tangible resilience metrics. For multi-agent systems,  maximizing system
resilience hinges on the ability of agents to judiciously decide when and what to communicate (STL formulae, their quantitative semantics or representations thereof) based on their belief states, history and local knowledge.  
 In a similar vein, in multi-agent Kripke systems (modal logic),   the meaning of a logical formula (Kripke semantics) is defined in terms of a set of possible worlds, and a truth assignment assigning a truth value to each formula at each possible world.  This  allows agents  to reason about the truth of a formula in different possible worlds and modeling other agents' knowledge  to refine their understanding of the world for efficient collaborative problem-solving.  
 Finally,  owing to the fact that cyber-physical systems typically consist of multiple interconnected components and operate in high-dimensional and continuous state spaces, the ability to decompose these systems into verifiable subsystems, each adhering to local constraints offers a scalable and efficient approach for  compositional formal verification.

\subsection{Emergence} 

\begin{tcolorbox}[colback=yellow!10, colframe=white!20!black, boxrule=1.9pt, boxsep=2pt,left=2pt,right=2pt,top=2pt,bottom=2pt]
    \noindent \textsf{Q4}: What is the impact of network topologies (e.g., scale-free and small-world),  higher-order interactions and their dynamics on the emergence of resilience? What are the scaling laws of resilience in terms of node density, connectivity and topological structures?
    \end{tcolorbox}

Complex networks, ranging from transportation systems to power grids and wireless networks, are dynamic structures characterized by heterogeneity, modularity and hierarchy. These structures have profound implications for how information is processed,  composed, and how systems build resilience and adapt to perturbations at various spatiotemporal scales. Due to their highly heterogeneous connectivity, many real-world networks exhibit robustness to random failures (e.g., node or edge failure).     In contrast to regular networks (such as grids or lattices),  heterogeneity introduces a diversity of possible failure scenarios and attacks. While homogeneous networks respond similarly to perturbations—owing to uniform node degrees—heterogeneous networks such as scale-free networks exhibit non-uniform responses, where failures affecting high-degree nodes (hubs) can cause catastrophic collapses, whereas the same perturbations is  absorbed in homogeneous settings.
\begin{wrapfigure}{r}{0.5\textwidth} 
    \centering
    \vspace{-1pt}
        \includegraphics[width=\linewidth]{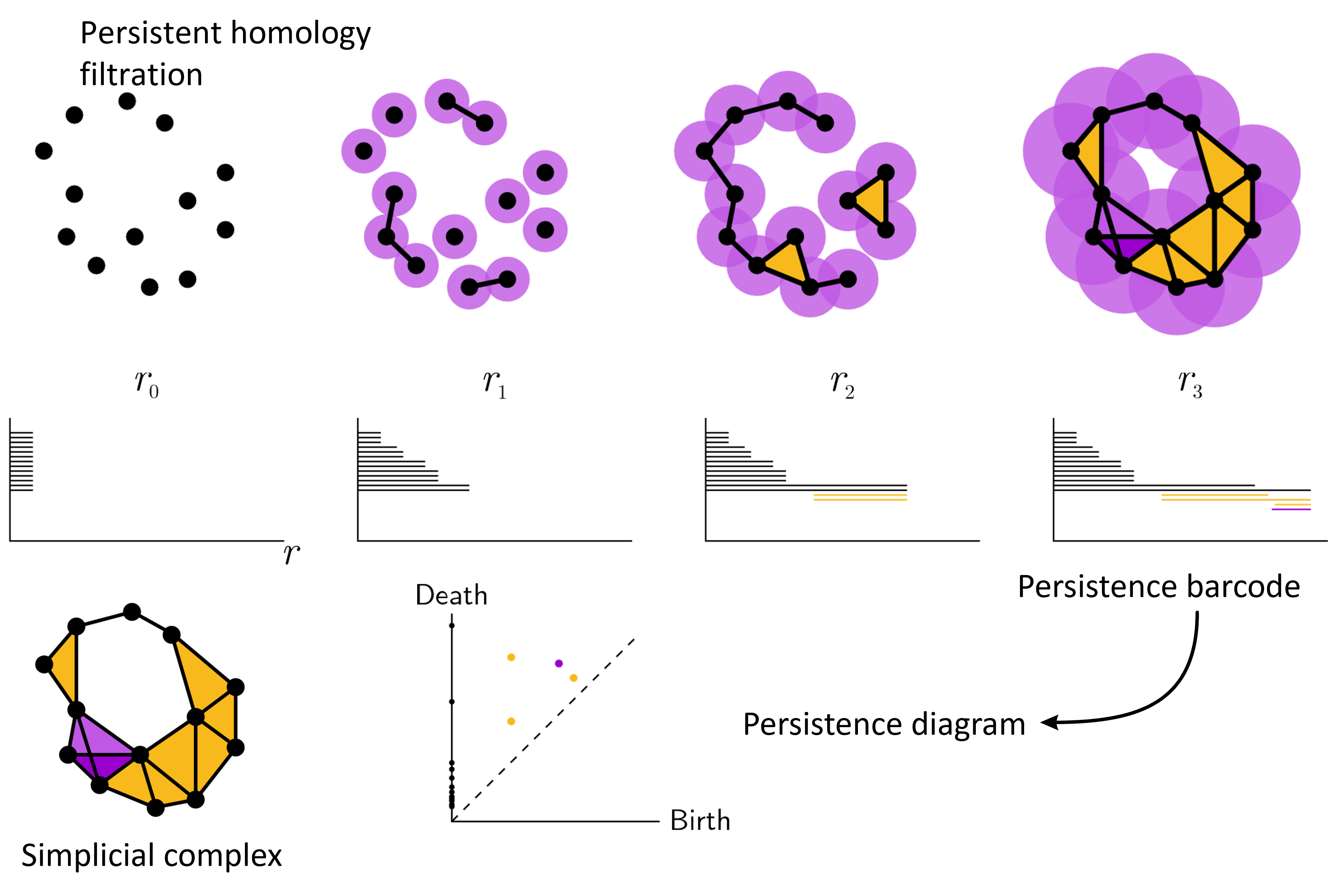} 
    \caption{\footnotesize Simplicial complexes, persistent homological filtration and barcodes as topological abstractors.}
    \label{fig561}
    \vspace{-2pt}
\end{wrapfigure} 
Understanding how different network topologies, scale-free, small-world, hierarchical, etc., respond to disruptions is critical. Traditional global metrics (e.g., node degree distribution, mean degree, or betweenness centrality) offer a coarse understanding. In contrast, higher-order local structures, such as network motifs \cite{dey2019network, doi:10.1126/science.298.5594.824}), offer a finer-grained geometric and functional description. Motif-based analysis focuses on the distribution, concentration, and lifetime of multivariate subgraph patterns and how they contribute to the network's global resilience. Viewed through the lens of algebraic compositionality, resilient systems can be constructed by composing the reliability of their individual motifs. This approach moves beyond mean-field or i.i.d. assumptions and emphasizes  interdependencies among subsystems, enabling more accurate assessments of how local failures percolate through the system. To rigorously model such interdependencies, copulas provide a powerful statistical framework. By decoupling the marginal distributions of individual agents from their dependence structure, copulas overcome the limitations of Gaussian or multivariate t-distributions \cite{copula}. This is especially important in multi-agent systems, where coordinated behavior emerges from complex interactions. Copula-based approaches involve three steps: selecting marginal distributions, specifying a copula function, and estimating parameters with model validation. 
Extending beyond graph-theoretic methods, topological data analysis (TDA) offers insights into the higher-order relational geometry of networks. In particular, persistent homology extracts topological invariants—such as loops or voids—through a multi-resolution analysis of simplicial complexes (known as \emph{filtration}). TDA metrics like Betti numbers characterize the birth and death of homological features over varying scales (e.g., a spatial or temporal radius ($r$ in Fig. \ref{fig561})), providing a topological abstraction  and  nuanced  resilience metric \cite{asirimath2024rawdatastructuralsemantics}.   

Besides these topological aspects emphasizing the importance of structure, a dynamical systems perspective pertaining to the   structural stability of the attractor landscape is crucial for resilience. Depending on the perturbation magnitude and (low/high)-order of interaction, as a node/agent or subsystem approaches a bifurcation point, it may lose resilience and shift abruptly to a different attractor.  Bifurcations capture qualitative changes in system behavior and are linked to the structure and stability of attractors, undergoing transformation at bifurcation points.  
Analyzing the bifurcation structure helps anticipate how the system responds to changes  and enables the design of adaptive mechanisms to steer the system away from undesirable operating regimes.   Specifically,  when the system’s configuration or parameters are far from a bifurcation point, the basin of attraction tends to be large, yielding
a faster rate of recovery. Conversely, as the system nears it, the basin of attraction contracts, yielding slower recovery from even small perturbations. This highlights the importance of the basin size as a  resilience metric, particularly  in terms of recoverability. 

\begin{figure}\centering 
	\includegraphics[width=8cm]{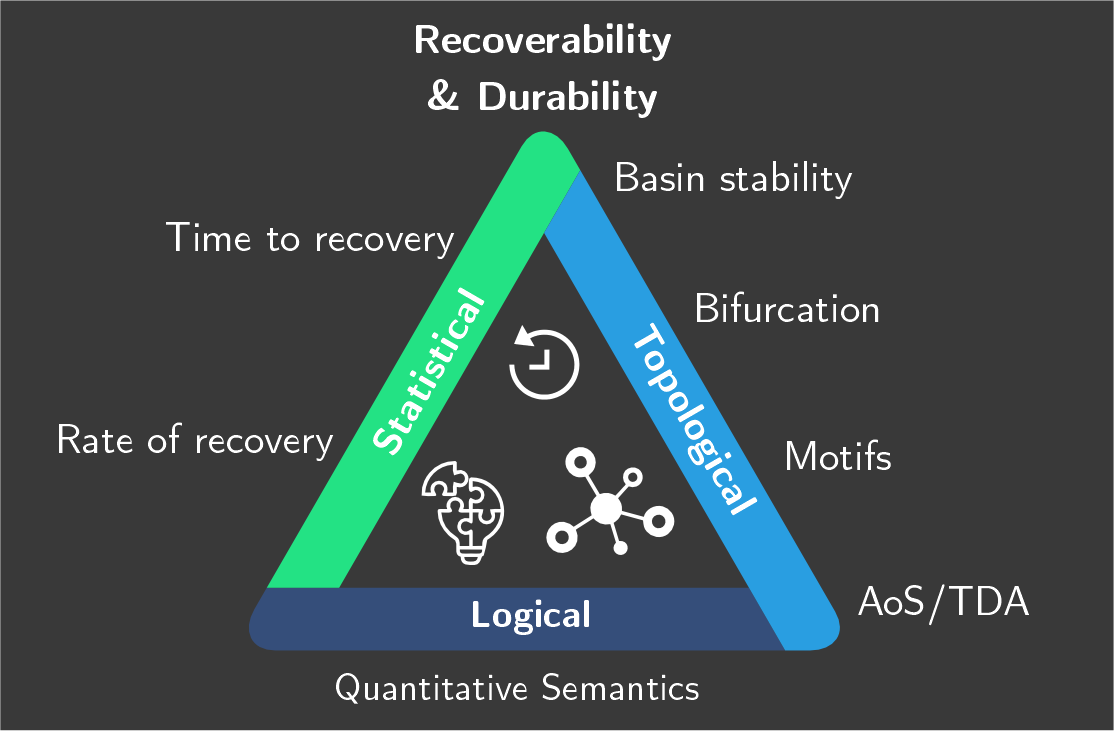}         	\caption{ \footnotesize Resilience metrics along the statistical, topological and logical continuum.}
    \label{metrics}
\vspace{-3pt}
	\label{fig17}
\end{figure}

 \vspace{-1mm}
\section{Metrics and Tradeoffs} 
\subsection{Metrics}

``\textit{If we cannot measure or quantify resilience, we should not discuss it}.'' This statement underlines the necessity for clear metrics when evaluating resilience. Key primary metrics include: (i)   Resistance (Inertia): the system's ability to withstand perturbations without performance degradation; (ii)   Recoverability: the time required to return to a baseline or acceptable state after disruption;  (iii)     Durability: the degree to which functionality is preserved, whether through elastic (reversible) or plastic (adaptive) responses. The relevance of these resilience metrics depends on the type of sensory data structures and potential use cases.  It is also worth emphasizing that resilience metrics include metrics for detecting/inferring node and network resilience, as well as metrics for dealing with recoverability and durability (post-disruption).
 From a \textbf{topological perspective}, resilience refers to an agent’s or network’s capacity to maintain its intrinsic geometric or topological structures under perturbations. Here,  TDA provides efficient tools to uncover such invariants by analyzing data across multiple resolutions via nested simplicial complexes. These invariants serve as “structural X-rays” revealing essential shapes and patterns. Important topological metrics include:
\begin{itemize}
    \item 
    Persistence diagrams (PDs), which capture the birth and death of topological features.

\item     Age of Structural Semantics (AoS), quantifying the persistence of internal structures over time.

\item     Motifs, which are recurring higher-order local substructures in networks.

\end{itemize}

\noindent From a dynamical systems viewpoint, resilience can be assessed through the lens of:
\begin{itemize}
    \item     Linear stability, which reflects the response to small disturbances and analyzes how a system behaves near an equilibrium point.

\item    Basin stability, which quantifies how resilient a stable state is by measuring how large the basin of attraction is, capturing robustness/resilience against large perturbations.

\item     Distance to bifurcation, indicating proximity to critical points where  system behavior changes.

\end{itemize}

\noindent For safety/mission-critical systems, statistical measures are  inadequate. Instead, formal logic frameworks, such as Signal Temporal Logic (STL) and modal logics, offer guarantees by specifying resilience properties syntactically (via formulas) and semantically (via recoverability and durability pairs  satisfying the STL formula). This approach enables the rigorous specification and verification of resilience. Last but not least,  the compositional techniques (discussed in Q2) provide a rich calculus allowing to compare and compose resilient systems. For example, one network may be  resilient locally but not globally. On the flipside, a globally resilient system may emerge from the integration of non-resilient components.  

\noindent A summary of resilience metrics across different frameworks is given below:

\begin{itemize}
    \item     \underline{Topological}: Concentration and lifespan of motifs, persistence diagrams (e.g., Betti numbers), and measures like AoS or topological basin analysis.

 \item    \underline{Dynamical}: Linear and basin stability, size and geometry of attractor basins, and distance to bifurcation.

\item     \underline{Logical (temporal, modal)}: Quantitative semantics using recoverability/durability pairs, signal distance from STL-satisfying trajectories, soundness and completeness of formal specifications.

\item     \underline{Statistical}: Joint probabilities of disruption and bounded degradation, recovery rate and time of recovery, area-under-curve metrics and metaresilience \cite{kim2025resilientllmempoweredsemanticmac}. In distributed consensus problems, (r,s)-robustness (a form of Byzantine resilience) provides sufficient conditions for consensus under $r$ adversarial agents \cite{CivitDGGKV24}.
 
\end{itemize}

\vspace{-2mm}

\subsection{Selected Tradeoffs}
Resilience involves trade-offs at both node (local) and system  (global) levels, spanning several key dimensions:
\begin{itemize}
\item     \underline{Recoverability vs. Durability}: Is it more beneficial for a system to quickly recover from disruptions yet have lower durability, or have longer durability  albeit slower recovery (see Fig. 2)?
    \item    \underline{Energy vs. Resilience}: To what extent does enhancing system resilience necessitate increased energy expenditure? This trade-off is particularly important in resource-constrained environments where efficiency is critical.

\item     \underline{Resilience–Robustness–Efficiency}: Given limited resources, how should a network strike a balance between preparing for unforeseen disruptions and maintaining operational efficiency after disruption?

\item     \underline{Connectivity-Topology-Dynamics}:  What are the critical thresholds in connectivity or topology needed to maintain system resilience? How does resilience scale with node density, higher-order interactions and network topology?   multi-connectivity, such as all-to-all links, can enhance resilience to node or link failures, but incurs higher energy consumption. In contrast, sparser connectivity may improve energy efficiency but reduce system resilience. Moreover, higher-order interactions  play a key  role in improving adaptation and resilience via linear stability (resistance to small disturbances), albeit  shrinking the basin of attraction (robustness to large perturbations), making  them harder to detect \cite{krakov}. The basin stability, a global measure based on the size of these attraction basins, determines the system’s resilience to large perturbations. 
\item     \underline{Resilience-Sustainability}:  Resilience is   closely related to sustainability, though the two are not synonymous. Indeed, sustainability implies resilience, but the reverse is not always true. Brundtland  \cite{wced1987common} defined sustainability ``as meeting present needs without compromising the ability of future generations to meet theirs'', underscoring the need for   efficient resource usage to ensure survival in a constantly changing environment. While resilience at the  agent level is essential for immediate survival, it is not sufficient for long-term sustainability. For instance, rapid adaptation to sudden disruptions may help an individual persist in the short term, but without strategic planning, such responses can  yield unsustainable outcomes. Achieving sustainability therefore demands proactive planning and preparedness over extended time horizons. At the agent level (locally), this requires minimizing the expected free energy  by seeking optimal actions that balance risk and uncertainty. At the system level, sustainability requires the alignment of multi-agent interactions, such that each agent’s optimal policy also contributes to reducing the collective system utility. Ultimately, this means promoting a stable state across different spatial and temporal scales, where resilience at each hierarchical level supports resilience throughout the system. Such systemic alignment enables coordinated actions and adaptations over time, reinforcing both resilience and long-term viability in harmony with the environment.
\end{itemize}

\section{Conclusions }
Despite its critical importance, resilience, a foundational concept in 6G, remains underexplored, often conflated with robustness and lacking well-defined metrics. Yet, resilience holds significant potential to advance and refine current incremental 6G visions.   In contrast to robustness and reliability, resilience assumes  that
disruptions will inevitably happen. Resilience, in terms of elasticity, focuses on the ability to
bounce back to favorable states, while resilience as plasticity involves agents (or networks) that can
flexibly expand their states, hypotheses and course of actions, by transforming through real-time
adaptation and reconfiguration. This constant situational awareness and vigilance of adapting world
models and counterfactually reasoning about potential system failures and best
responses, is a core aspect of resilience. By embedding resilience into  networks, ecosystems, social systems, and economic infrastructures, we can improve adaptability, mitigate failure risks and cultivate sustainable future-ready systems.

This article has just scratched the surface. Future work will delve into the details of the principles of resilience and, more importantly, showcase its relevance and importance in concrete real-world  scenarios.

\bibliographystyle{ieeetr}  
\bibliography{IEEEabrv,resilience} 
\end{document}